# Acceleration of galactic electrons at the solar wind termination shock and Voyager 1 observations

**Marius S Potgieter**[1]**, P. Lourens Prinsloo, R du Toit Strauss**
*Centre for Space Research, North-West University, 2520 Potchefstroom, South Africa*
*E-mail:* marius.potgieter@nwu.ac.za

Diffusive shock acceleration (DSA), as an acceleration process for Galactic electrons at the solar wind termination shock (TS), is investigated with a comprehensive numerical model which incorporates shock-acceleration, particle drifts and other major modulation processes in the heliosphere. It is known from our previous work that the efficiency of DSA depends on the shape of electron spectra incident on the TS, which in turn depends on the spectral shape of the very local interstellar spectrum. Modulation processes also influence the efficiency of DSA. We find that TS accelerated electrons can make contributions throughout the heliosphere to intensity levels, especially at lower energies. An interesting result is that increases caused by DSA at the TS are comparable in magnitude to electron intensity enhancements observed by Voyager 1 ahead of the TS crossing. These intensity increases are large enough to account for observed intensity peaks, and thus supports the view that DSA is involved in their production. Additionally, the energy spectra observed by Voyager 1 throughout the heliosheath are reproduced satisfactorily, as well as the PAMELA spectrum at Earth at higher energies. We also find that an increase in the rigidity dependence of the diffusion coefficients for these low-energy electrons seems required to reproduce the spectral shape of observed modulated spectra in the heliosheath at kinetic energy $E < \sim 4$ MeV. This is different from intermediate energies where rigidity independent diffusion explains observations satisfactory.

[1]Speaker

## 1. Introduction

Voyager 1 & 2 crossed the solar wind termination shock (TS) at radial distances of ~ 94 AU (2004) and ~ 84 AU (2007), respectively [1,2]. Before this TS crossing by Voyager 1 (V1), two distinct enhancements of 6-14 MeV (or 5-12 MeV) electrons were observed, each lasting for several months [3,4]. They were interpreted as precursors to the TS crossing, probably as the result of the streaming of accelerated particles along short-term magnetic field connections between the TS and the spacecraft [e.g. 5]. The highest level of these intensity enhancements is significantly larger than the background intensity, by a factor of ~ 2.5. These particle events are considered to be indicative of the acceleration of Galactic electrons at the TS, conceivably caused by diffusive shock acceleration (DSA). The latter has come under much scrutiny when it was discovered that the acceleration region of anomalous cosmic rays (ACRs) was not at the location of the TS [1,6]. The hypothesis of ACRs being formed by DSA of pick-up ions at the TS became questioned and interest in finding a substitute explanation subsequently shifted to alternative acceleration mechanisms; for a review of these mechanisms, see [7] and references therein. However, some authors maintain that DSA remains a viable mechanism, while the power-law distributions of TS particles observed below 3 MeV nuc$^{-1}$ [1] are still considered to follow from DSA. Although the involvement of DSA is questioned in the acceleration of very-low-energy ions there has not been a formal investigation into its viability as an acceleration mechanism for Galactic electrons at the TS; see also the related discussions by [8].

A few important features of electron modulation are to be emphasized, e.g., when V1 crossed the heliopause (HP), it was revealed [10,11] that the very local interstellar spectrum (VLIS) of these electrons is power-law distributed at kinetic energies 6–60 MeV [12,13] while the very large intensity gradient observed across the heliosheath [14] is indicative of efficient particle scattering in this region. These aspects are illustrated and elaborated on below.

## 2. The re-acceleration model for electron modulation

This DSA-drift-numerical model was described in detail by [9], who also motivated the assumptions for the diffusion coefficients (DCs) and modulation parameters such as the solar wind profile and the heliospheric magnetic field (HMF). It suffices to say that the location of the TS is specified at $r_{TS}$ = 94 AU, and the HP at $r_{HP}$ = 122 AU, with a heliosheath in between. The compression ratio of the TS was assumed to be $s$ =2.5. The VLIS for electrons is according to [16; see also 17]. The assumed set of modulation parameters in the model reproduces electron spectra observed by PAMELA [19,20] at the Earth as a validation check.

DSA effects are studied for the two drift cycles based on the HMF polarity orientation. All numerical solutions are shown along the line of travel of V1 at a polar angle of $\theta$ = 55°, at the Earth with $\theta$ = 90°, and for solar minimum conditions with the heliospheric current sheet (HCS) tilt angle $\alpha$ = 10°. For the illustrative solutions shown below, DSA is switched on and off in the model, while the drift coefficient, the DCs, also $s$, $r_{TS,}$ and $r_{HP}$ are kept unchanged, except for changing the polarity of the HMF (so-called A < 0 and A > 0 cycles). Because CR modulation beyond the HP is quite small [e.g. 18], the HP is assumed as the outer modulation boundary. Elaborate discussions of the applied model, its solutions from the HP through the heliosheath and across the TS to the Earth, its implications, and conclusions can be found in [9,15].

## 3. Results and discussion

The solutions from the model to illustrate the effects of DSA on the radial intensity profiles for electrons of varying kinetic energy *E* are shown first. In Fig. 1, it is done at 16, 200 and 1000 MeV, and for the two HMF polarity cycles; A < 0 (top panel) and A > 0 (lower panel). The solid lines represent profiles with DSA occurring at the TS, compared to intensity profiles without DSA. First, notice how these profiles differ from A < 0 to A > 0, as an illustration of drift effects. Obviously, the associated radial gradients are much larger for the lower energies. The important issue here is how much the solutions differ from those with and without DSA. A noticeable feature is the very large intensity gradients beyond the TS in order to match the value of the VLIS at these energies at the HP. Evidently, across and beyond the TS the consequence of DSA on electron modulation is to reduce these gradients but these effects are progressively over-taken by the sharp increase of Galactic electrons towards the HP.

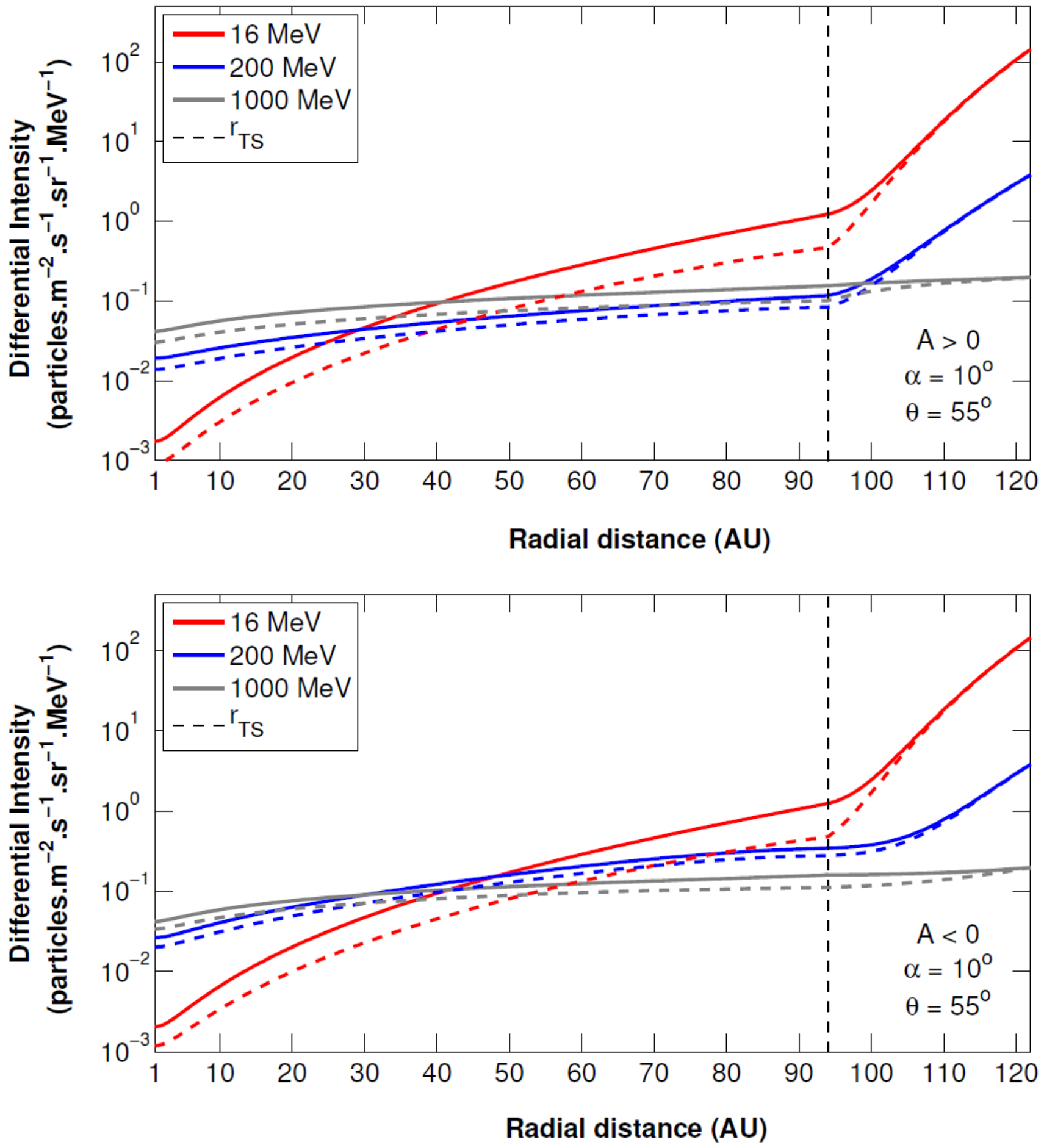

**Figure 1:** Radial intensity profiles for electrons with kinetic energies of 16, 200 and 1000 MeV, for A > 0 (top panel) and A < 0 (lower panel) polarity cycles. Profiles with DSA (all solid lines) at the TS ($r_{TS}$ = 94 AU; vertical dashed line) and without DSA (all dashed lines) are shown. The VLIS is specified at the HP ($r_{HP}$ = 122 AU). Profiles are shown for a polar angle of $\theta$ = 55° (V1's trajectory) and the HCS tilt angle $\alpha$ = 10°, representing solar minimum conditions.

In Fig. 2, the corresponding ratios of the radial profiles shown in Fig. 1 are displayed as an illustration of the 'magnitude' of the contribution by DSA to electron intensities at the TS (profiles with and without DSA). The solid lines are for A < 0 and the dashed lines are for A > 0 HMF polarity. At 16 MeV this ratio is as large as a factor 2.5. Surprisingly, this factor is almost 1.5 for 1000 MeV electrons, depending on the drift cycle, and somewhat larger than for 200 MeV electrons. The reason why this happens is quite interesting and is explained in detail by [9]; basically, it depends on the spectral shape of the electron VLIS.

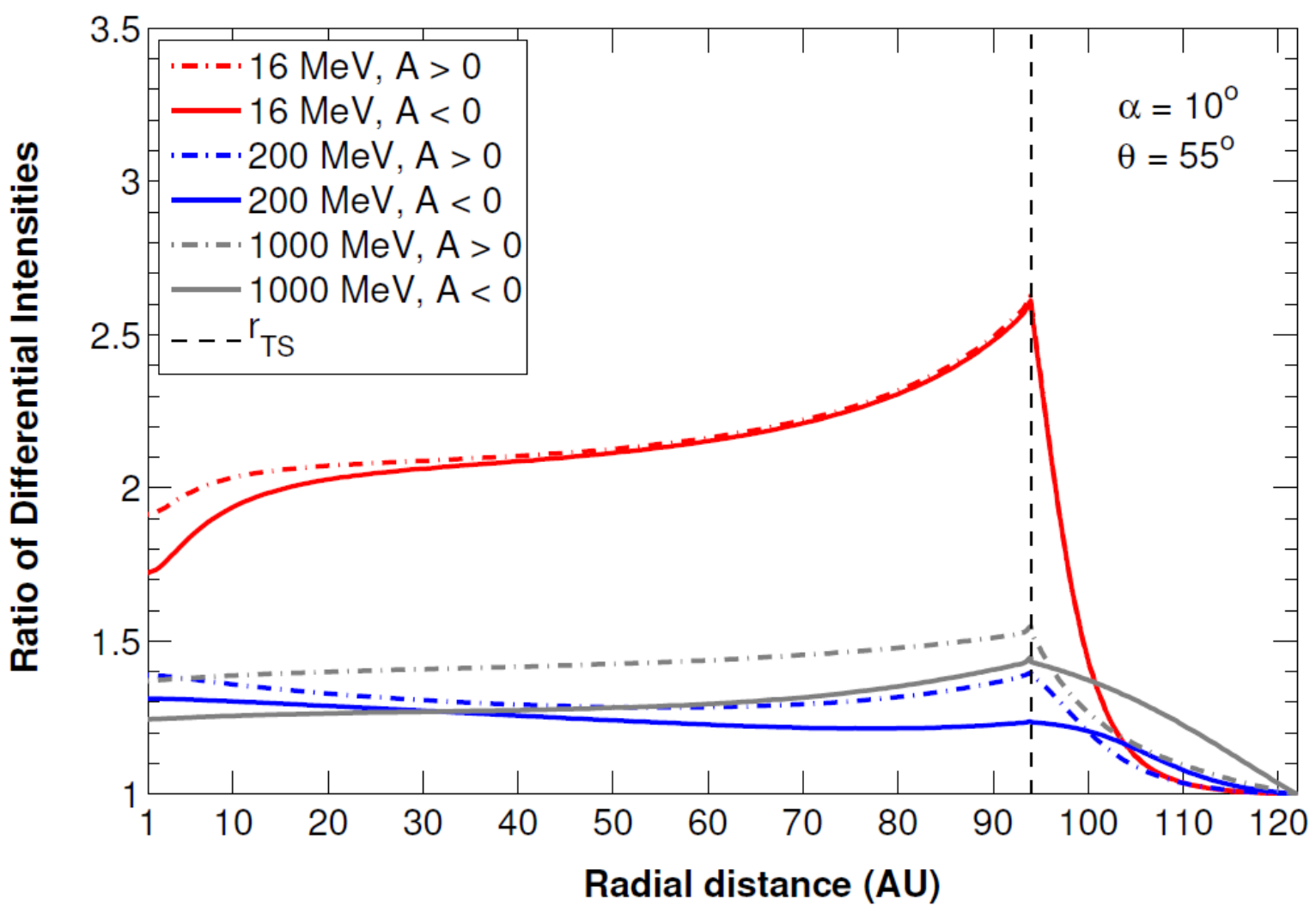

**Figure 2:** Ratios of the radial profiles of Fig. 1, with DSA at the TS to those without DSA, shown for the A < 0 (coloured, solid lines) and the A > 0 HMF polarity cycles (coloured dashed-dotted lines) at the indicated energies. This ratio is indicative of the increase in intensity that DSA causes at the TS and how this is distributed throughout the heliosphere. Note the drift effects (differences between A < 0 and A > 0 cycles) and especially how sharply the DSA contribution drops off beyond the TS. See also [9].

In Fig. 3 and Fig. 4, the numerical solutions in the form of energy spectra and radial profiles are compared to available observations. Fig. 3 shows modelled spectra at the Earth, at the TS and into the heliosheath, compared to V1 electron measurements for ~4-40 MeV [Webber, private comm. 2012] and from PAMELA at higher energies [19]. We note the following: (1) The modelled spectra demonstrate how the power-law distribution of the VLIS, which provides a satisfactory fit to observations at 122 AU, is mostly retained in the heliosheath despite the significant modulation. The V1 measurements also clearly reflect this. (2) The lowest-energy measurements, at ~4 to 8 MeV, increasingly deviate with decreasing radial distances from this power-law shape. (3) The spectrum without DSA at 94 AU does not reproduce the observational data, while the TS spectrum with DSA does quite well. But, these DSA effects are decimated with increasing radial distance, as shown in Fig. 1 and 2. This means apart from DSA, another modulation effect must play a role. (4) The latter may indicate that the assumption of rigidity-independent diffusion for electrons may only be applicable down to these energies and that below these energies, the DCs (corresponding mean free paths MFPs) increase and thereby decrease modulation. The required MFP up-turn is initiated at lower rigidities at 94 and 118 AU than at the distances in between. The V1 measurements also suggest that the underlying turbulence displays a radial dependence, since the MFP up-turn at the lowest considered energies becomes less pronounced with increasing distance from the TS. (5) Also shown is that modelled spectra at Earth closely follow the PAMELA measurements [19] at $E >$ ~100 MeV. This confirms that in addition to the correct levels of modulation achieved in the heliosheath, the total global modulation, between the HP and the Earth, at these energies is also accurately reproduced. (6) The shape of the VLIS below 1 MeV is speculation because it is unknown what the VLIS for electrons is below 5 MeV, and needs to be investigated further. We consider these displayed values as far too high for $E < 1$ MeV.

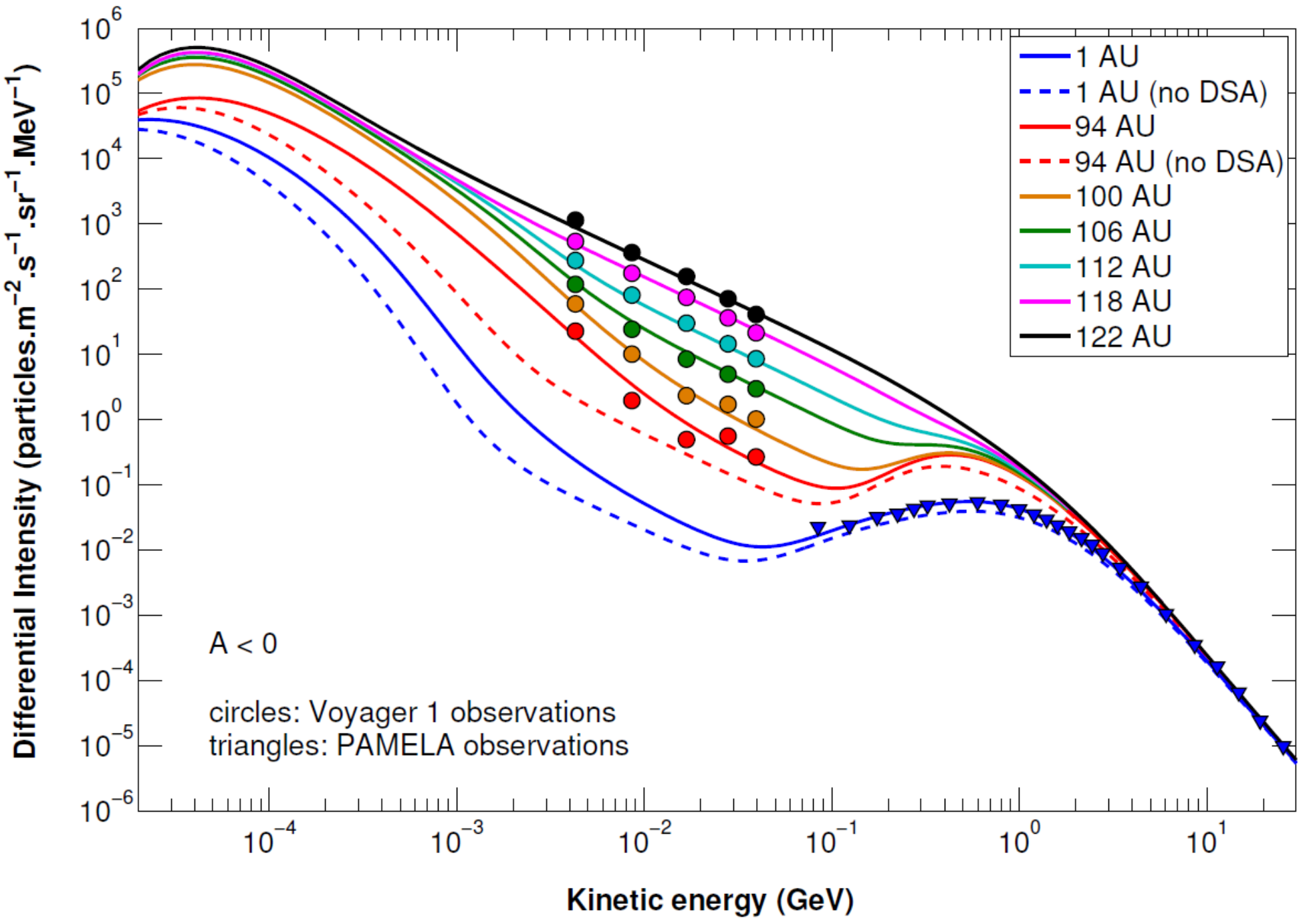

**Figure 3:** Modelled spectra for Galactic electrons at $\theta$ = 55°, during an A < 0 cycle, are compared with observations at and beyond the TS (94 AU to 118 AU) from V1 (coloured, filled circles), and from PAMELA (triangles) at the Earth during 2009 [19,20]. Varying colours represent different radial distances, from 1 AU (Earth; $\theta$ = 90°), to 94 AU where the TS is located, to 122 AU where the HP is located. The VLIS is shown as a solid black line at 122 AU. Solid and dashed lines represent solutions with and without DSA effects, respectively. Note the differences at 1 AU and 94 AU and that it dissipates quickly beyond the TS. The shape of the VLIS below 1 MeV is conjecture [15], and considered to be far too high.

In order to reproduce the observed intensities for 5-12 MeV electrons [Webber, private comm. 2014] in the heliosheath, varying modulation conditions had to be simulated in the model throughout this region. At least five diffusively constrained regions for the heliosheath had to be assumed. The subsequent profiles, with and without DSA, are shown in Fig. 4 at a representative energy of 6 MeV together with the observations from V1. This approach reproduces the observed radial distribution of intensities adequately. From the TS outwards to at least 100 AU, the effects of DSA are essential to account for the shape and trends of the increasing intensities, while modelled and observed intensities are also similar for $r > 109$ AU, where DSA effects are weakened. At intermediate distances (103-112 AU), the observations are better represented without DSA effects; the accelerated contribution causes the model to overestimate intensities in this region. At $r < r_{TS}$, the model predicts intensities at about the same level of the highest points of the observed intensity peaks. The factor of ~ 6 by which these intensities are increased from levels without DSA is significant, however, evidently larger than the magnitude of the two observed peaks, but here displayed with respect to the observed background levels (red symbols), which is considered artificially high because of proton contamination in the V1 detector as discussion by [21].

It is concluded that DSA as implemented in the model is efficient enough to produce intensity increases of the magnitude of these peaks. The quantitative features of electron DSA presented here are of course subject to assumptions made for the transport coefficients; smaller DCs would enhance the effects while larger DCs would diminish it; see the discussions by [9], and particularly [22,23] who illustrated DSA effects for electrons, positrons, protons and anti-protons at the TS.

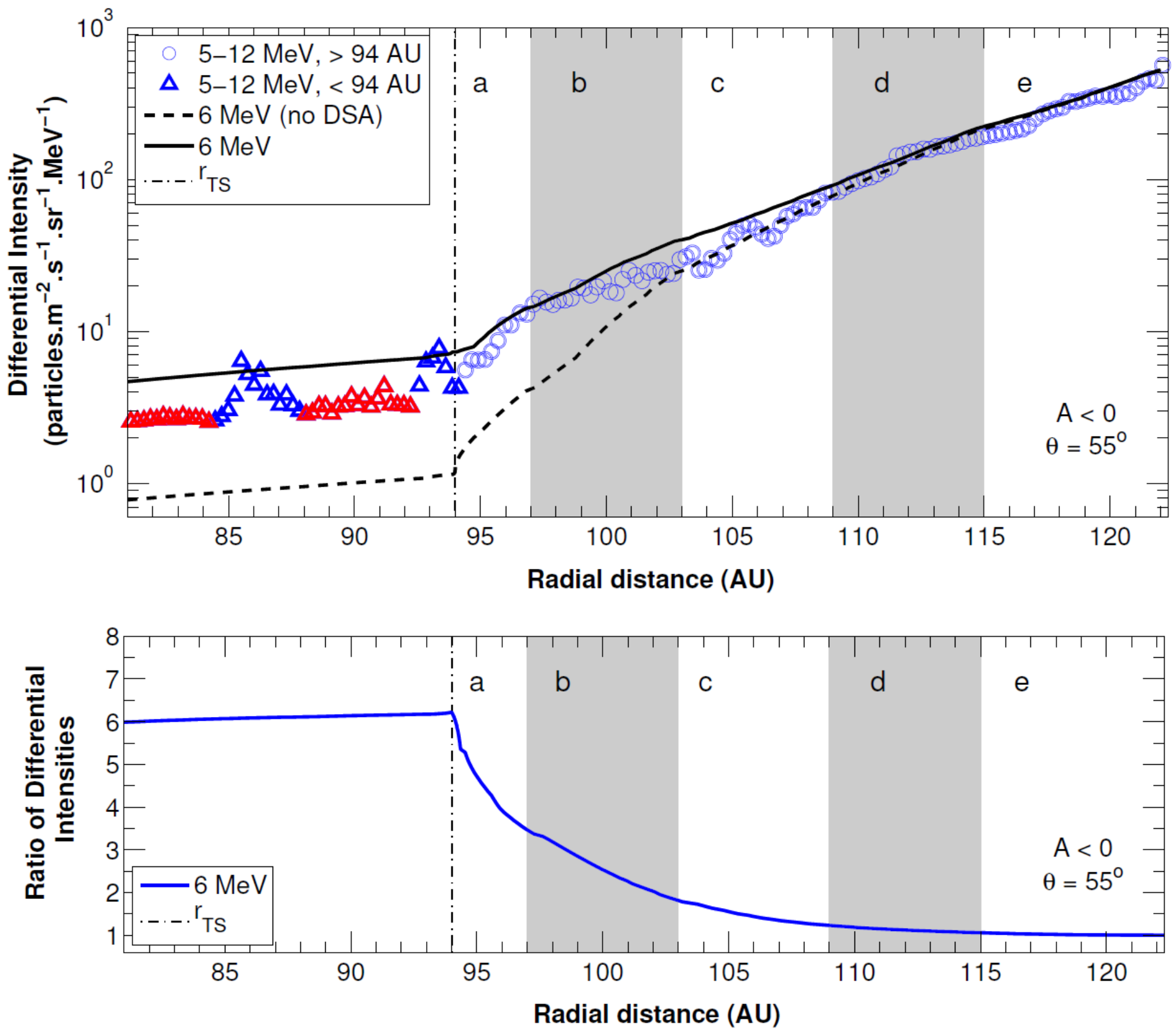

**Figure 4:** Computed radial profiles, with DSA (solid black line) and without DSA (dashed black line) for 6 MeV electrons based on combining five segments with differently assumed DCs in each region, labelled a to e. V1 observations between 5-12 MeV, for $r < r_{TS}$ and $r > r_{TS}$, are given by triangular and circular symbols, respectively [14; Webber, private comm.]; data points in red represents background intensities, which could be artificially high because of proton contamination in the V1 detector. The bottom panel shows the corresponding ratio of the computed profiles with and without DSA from the upper panel. This is done for an A < 0 polarity cycle and with polar angle $\theta = 55°$ in the model.

## 5. Conclusions

When V1 approached the TS at 94 AU, the electron detector on board measured two distinct enhancements of 5-12 MeV electrons, with intensity far higher than the background level, each lasting for several months. These particle events are thought to be indicative of the acceleration of Galactic electrons at the TS, conceivably caused by diffusive shock acceleration (DSA). This is extensively investigated with a DSA-drift-modulation model [see 9].

We found: (1) DSA can increase electron intensities at 6 MeV at the TS by a factor of ~ 6, at 16 MeV by a factor of ~ 2.5, and surprisingly by a factor of ~ 1.6 at 1 GeV, but dissipating quickly beyond this energy. This is possible because of the particular shape of the VLIS for electrons which is the source of electrons to be accelerated at the TS; the efficiency of DSA depends strongly on the shape of the modulated electron spectrum incident at the TS, which in turn depends on the features of the VLIS; see also e.g. Fig. 2 as published by [9].
(2) Because of solar modulation, these factors reduce to between 1.7 and 1.9 at the Earth for 16 MeV, and between 1.2 and 1.4 at 1 GeV, depending on the drift cycle. DSA electrons can thus make an appreciable contribution to the intensities of Galactic electrons even at the Earth.
(3) Particle drifts [24,25] are found to play an influential role in the DSA process at higher energies, affecting both the magnitude of DSA-associated increases at the TS and the transport of accelerated electrons throughout the heliosphere.
(4) In the heliosheath DSA effects subside within ~15 AU from the TS. Because the VLIS for electrons is quite high at 6 MeV, the increased intensities produced by DSA at the TS is no match for the incoming Galactic electrons at a given energy.
(5) In order to simulate the extra-ordinary increase in low-energy electrons inside the heliosheath, this region had to be divided in five distinct diffusive regions, each with a peculiar set of propagation parameters; see discussions by [14,15 and their Table 7.1].
(6) It seems unlikely that the acceleration of these low-energy electrons by DSA at the TS would make any significant contribution to electron intensity in the very local interstellar medium.
(7) The predicted increase of a factor of ~ 6 of 6 MeV electrons at the TS is enough to explain the observed transient type increases of 5-12 MeV electrons from V1. It is concluded that DSA as implemented in the model is efficient enough to produce intensity increases of the magnitude of these peaks near the TS.

The authors acknowledge the partial financial support of the South African National Research Foundation (NRF). MSP thanks WR Webber for many interesting and useful discussions on the V1 data and making the 4-40 MeV and later the 5-12 MeV electron data available for the Master's thesis of PLP.